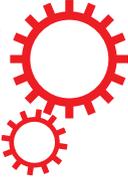

# OPEN

# One-Step Fabrication of pH-Responsive Membranes and Microcapsules through Interfacial H-Bond Polymer Complexation

Julien Dupré de Baubigny[1], Corentin Trégouët[1], Thomas Salez[2,3], Nadège Pantoustier[1], Patrick Perrin[1], Mathilde Reyssat[2] & Cécile Monteux[1,3]

Biocompatible microencapsulation is of widespread interest for the targeted delivery of active species in fields such as pharmaceuticals, cosmetics and agro-chemistry. Capsules obtained by the self-assembly of polymers at interfaces enable the combination of responsiveness to stimuli, biocompatibility and scaled up production. Here, we present a one-step method to produce *in situ* membranes at oil-water interfaces, based on the hydrogen bond complexation of polymers between H-bond acceptor and donor in the oil and aqueous phases, respectively. This robust process is realized through different methods, to obtain capsules of various sizes, from the micrometer scale using microfluidics or rotor-stator emulsification up to the centimeter scale using drop dripping. The polymer layer exhibits unique self-healing and pH-responsive properties. The membrane is viscoelastic at pH = 3, softens as pH is progressively raised, and eventually dissolves above pH = 6 to release the oil phase. This one-step method of preparation paves the way to the production of large quantities of functional capsules.

Non-covalent interactions are involved in complex self-assembled materials, which are found in nature[1] from DNA[2,3] to lipid membranes[4] and proteins[5]. Some of them involve hydrogen bonds[6], which have been intensively used to mimic nature and assemble supramolecular materials[7–11]. Recent years have seen increasing interest in the construction of coatings and membranes obtained by layer-by-layer (LbL) self-assembly of polymers on solid flat surfaces[12–14], colloidal particles[15] and even droplets[16,17]. Among all the studied systems, hydrogen-bonded polymer multilayers are of great interest, for instance for fuel-cell applications or proton-exchange flexible membranes[18]. Such multilayers also present unique stimuli-responsive properties by tuning pH, which allows for encapsulation and triggered release of active species at mild pH[19]. Recently, we showed that the interplay between hydrophobic interactions and hydrogen bonds allows to finely tune, by pH adjustment, the softening and breaking of a viscoelastic capsule membrane to eventually release inner compound[20]. However, the widespread development of such LbL membranes is limited by the time- and cost-consuming multiple-step fabrication process. Interfacial complexation involving oppositely-charged polymers has also shown great promise to achieve a one-step capsule production[21–23]. In these methods, ionizable polymers placed in oil and water phases assemble electrostatically at the oil-water interface to produce a thick and rigid membrane. In light of these studies, the one-step preparation of membrane by H-bond polymer complexation at a liquid interface could open a promising route to the large scale production of microcapsules as it overcomes the difficulties linked to a multi-step process.

In this Letter, we demonstrate the ease to make a micron-thick membrane at the interface between water and oil, through H-bonds polymer self-assembly. The membrane presents unique self-healing properties as polymers can diffuse from the oil and water reservoir phases to repair damages. We show that this system can be used to produce highly-stable capsules over a wide range of sizes, from tens of microns using microfluidics or a rotor-stator homogenizer, to the centimeter by dripping oil droplets into water. We successfully encapsulate droplets of biocompatible oils, Miglyol 812 (caprylic/capric triglyceride) and IPM (isopropyl myristate), with a membrane composed of biocompatible polymers, PMAA (poly(methacrylic acid)) and PPO (poly(propylene oxide)). Moreover, we show the pH-responsiveness of the membrane, as it softens when the pH is raised and even dissolves at pH > 6. Accordingly, the oil can be released from the capsules, thus offering new opportunities for encapsulation and triggered release of active species that could be potentially used in various domains such as environmental, cosmetic and pharmaceutical industries.

[1]ESPCI Paris, PSL Research University, CNRS UMR 7615, Laboratoire Sciences et Ingénierie de la Matière Molle, 10 rue Vauquelin, 75231, Paris, Cedex 05, France. [2]ESPCI Paris, PSL Research University, CNRS UMR 7083, Laboratoire Gulliver, 10 rue Vauquelin, 75231, Paris, Cedex 05, France. [3]Global Institution for Collaborative Research and Education, Global Station for Soft Matter, Hokkaido University, Sapporo, Hokkaido, 060-0808, Japan. Correspondence and requests for materials should be addressed to M.R. (email: mathilde.reyssat@espci.fr) or C.M. (email: cecile.monteux@espci.fr)

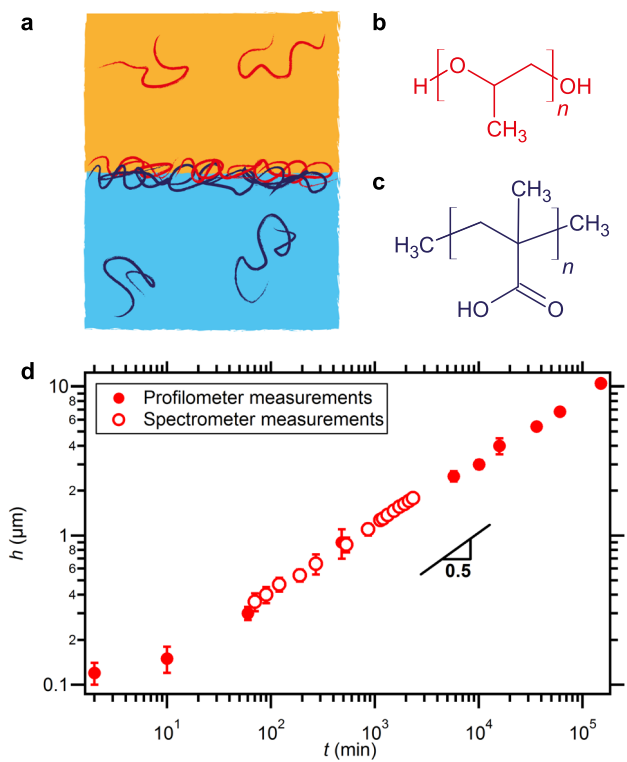

**Figure 1.** (**a**) Schematic of direct interfacial complexation between an oil phase (orange) and a water phase (blue). Each phase contains a polymer: (**b**) oil-soluble poly(propylene oxide) (PPO), and (**c**) water-soluble poly(methacrylic acid) (PMAA). A membrane is formed at the interface by H-bond complexation of the two polymers. (**d**) Evolution of the membrane thickness $h$ over time $t$. Measurements are performed by two methods: *in situ* spectrometry (empty circles) and *ex situ* profilometry (filled circles). Error bars are shown when larger than markers size.

## Results and Discussion

The membrane self-assembly is first probed in a model plane geometry. To prepare a flat polymer membrane, the aqueous phase containing the hydrogen-bond-donor polymer (PMAA) is put into contact with the oil phase containing the hydrogen-bond-acceptor polymer (PPO), as sketched in Fig. 1(a). The thickness of the membrane is measured either *in situ* using an optical spectroscope, or *ex situ* using an optical profilometer to analyze the membrane deposited from the liquid onto a glass slide. The membrane grows continuously with time without reaching any saturation over months and its thickness is 10 μm after 100 days as shown in Fig. 1(d). The thickness, which scales as $t^{1/2}$, strongly suggests a diffusion-limited mechanism. The associated diffusion coefficient can be determined from the extrapolated intercept of the long-time regime in Fig. 1(d). The obtained value is of the order of $10^{-17}$ m$^2$.s$^{-1}$ which is much lower than the diffusion coefficient of the polymers in the bulk phases. We therefore suggest that the growth of the membrane is controlled by the diffusion of the polymer molecules through the membrane to complex with their H-bond partner.

This model geometry further allows probing the self-healing properties of the flat membrane as shown in Fig. 2. The presence of the membrane at the oil-water interface is revealed by the formation of wrinkles as a gentle strain is applied with a spatula (Fig. 2(a)). Using a microscope cover slide, the initial polymer membrane is pushed on the side, which allows a fresh one to appear (Fig. 2(b)). A few seconds later, wrinkles can be seen at the oil-water interface as the membrane is gently pressed with a spatula again (Fig. 2(c)). This simple test shows that the membrane reassembles quickly from the oil and water reservoirs of polymer molecules. A double-wall-ring interfacial shear rheometer[24] is used to measure the surfacic shear storage and loss moduli of the membrane, $G'_s$ and $G''_s$, respectively, with increasing deformation amplitudes. Above a critical strain, the system is no longer in the linear regime and both moduli drop by several orders of magnitude. A visual inspection of the interface indicates that the membrane is damaged (Fig. 2 (d)). However, when the deformation is decreased again, the initial values of the moduli are recovered in about an hour. To explain these observations, we suggest that the membrane easily rearranges through diffusion of the polymer molecules towards and inside the membrane, owing to the H-bond non-covalent interactions.

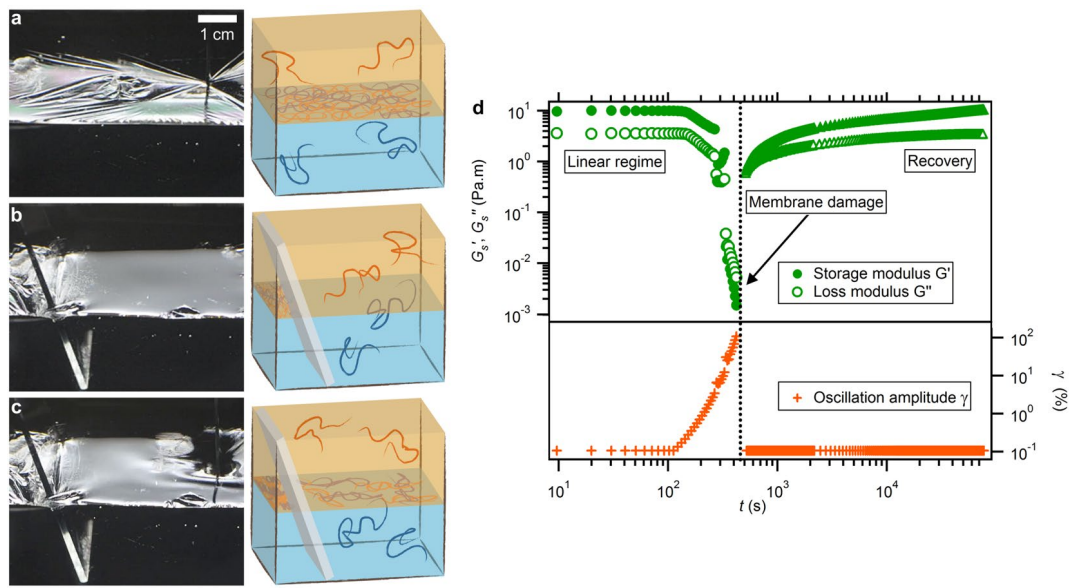

**Figure 2.** Self-healing process. (**a–c**) Polymer membranes formed at the oil (orange) - water (blue) interface. (**a**) Membrane formed during 15 hours for which a soft contact reveals a lot of wrinkles. (**b**) The membrane is destroyed and pushed side-way with a glass slide. (**c**) Wrinkles on the right of the image indicate that a new membrane is instantaneously formed. (**d**) Storage and loss moduli as a function of time: after a damage due to an increase of the oscillation amplitude, the membrane recovers its viscoelastic properties in the low-amplitude ("linear") regime. Error bars are shown when larger than markers size.

The polymer membrane can also self-assemble around oil droplets for encapsulation purposes. Capsules of various sizes can be obtained with this general principle, using three different methods. First, centimeter-sized capsules are easily produced by gently dripping oil drops containing PPO into a water phase containing PMAA (Fig. 3(a)). Secondly, using a rotor-stator homogenizer and by shearing the two fluid phases containing the polymers, capsules with a diameter of $25\,\mu m \pm 10\,\mu m$ are produced (Fig. 3(b)). Interestingly, this one-step process enables the preparation of microcapsules that are stable for months. Thirdly, our method is extended to microfluidics to produce a monodisperse population of micron-sized capsules. The microfluidic chip is composed of a flow-focusing unit where oil droplets containing PPO are produced in a pure aqueous phase to avoid the fast interfacial complexation at the constriction, which may plug the flow-focusing unit. The PMAA solution is then added right after to trigger the interfacial complexation at the oil-water interface. The capsules are finally collected in a chamber with a filter made by two-layer lithography[25], where they can be stored during several days (Fig. 3(c)). This semi-permeable filter being permeable to the external aqueous phase but not to capsules, allows to increase the capsule concentration. In these three systems, even in close contact to each other, all these capsules present a unique stability compared to standard surfactant-stabilized emulsions. It is especially the case for the centimeter-sized droplets which are known to be very difficult to stabilize. Indeed, the coalescence probability increasing with the droplet size, stable centimeter-sized emulsion droplets are scarce. Here, the high interfacial rigidity of the polymer membranes protects the assembly against coalescence during several days.

The rheological properties of the capsule membrane can be easily controlled using pH, which sets the number of protonated PMAA units. Frequency sweeps of the storage and loss surfacic moduli, $G_s'(\omega)$ and $G_s''(\omega)$, performed in the linear regime at the interface between the PMAA and the PPO solutions show that $G_s'(\omega)$ and $G_s''(\omega)$ vary over five orders of magnitude for pH ranging between 3 and 5.5, as shown in Fig. 4(a). At pH = 3, the membrane exhibits a viscoelastic behavior with a crossover frequency $\omega_c = 0.05 \pm 0.005\,\text{rad.s}^{-1}$. At low frequency ($\omega < \omega_c$), $G_s'$ is lower than $G_s''$ meaning that the layer presents a fluid behavior. At higher frequency ($\omega < \omega_c$), $G_s'$ is higher than $G_s''$ with $G_s'$ being almost independent of frequency, which is typically observed in polymer systems, and is known as the rubbery plateau[26]. The value of plateau modulus $G_s' = 7.1\,\text{N.m}^{-1}$ characterises the strength of the membrane. No widely spread method is used in the literature to measure such quantity and interfacial rheology has barely been used to investigate capsule membranes. In a previous article, dedicated to polymer multilayers at oil-water interfaces[20], we reported surface shear moduli of the order of $0.1\,\text{N.m}^{-1}$, which are one to two orders of magnitude lower than the results presented here. This could be due to the huge thickness difference between thin multilayers and our thick membrane. Other techniques in the literature including osmotic buckling[27], or AFM measurements[28], report Young modulus of the order of 1 to 100 MPa. In our experiment, if interfacial shear modulus is divided by the membrane thickness (540 nm), a bulk modulus $G' = 13.1$ MPa is obtained, which is on the same order of magnitude as what is found in the literature.

In bulk solutions of associative polymers, the crossover frequency would correspond to the relaxation time $\tau_c = 1/\omega_c$ required for chains to dissociate and flow. Similarly, our system can be viewed as an associative polymer membrane where PMAA and PPO chains stick together through hydrogen bonds and hydrophobic interactions. When the pH is raised from 3 to 5.1, the crossover frequency shifts to higher values, meaning that the chains dissociate faster. Indeed the number of hydrogen bonds between the two polymers becomes lower as the number of ionized PMAA monomer units increases. Remarkably, at an intermediate value of pH = 5.1 no rubbery plateau is observed in the explored frequency range, and the $G'_s(\omega)$ and $G''_s(\omega)$ curves remain parallel over 3 decades of frequency with $G'_s(\omega) \sim G''_s(\omega) \sim \omega^{0.7 \pm 0.05}$. This means that the system exhibits a distribution of relaxation times. Such a behavior has been predicted[29], and observed[30], for associative ionized polymer solutions with hydrophobic stickers. In our case, at intermediate pH, the chains are partly charged and the observed behavior could be due to hydrophobic interactions between PPO and PMMA. This complex behavior is currently under investigation. At pH = 5.5, the rheological response is purely viscous. Accordingly, at this pH no membrane could be detected using the optical spectroscopic technique. These observations reveal that the polymers do not self-assemble at the interface. Therefore the large number of PMAA charged units prevents efficient H-bonding and hydrophobic interactions.

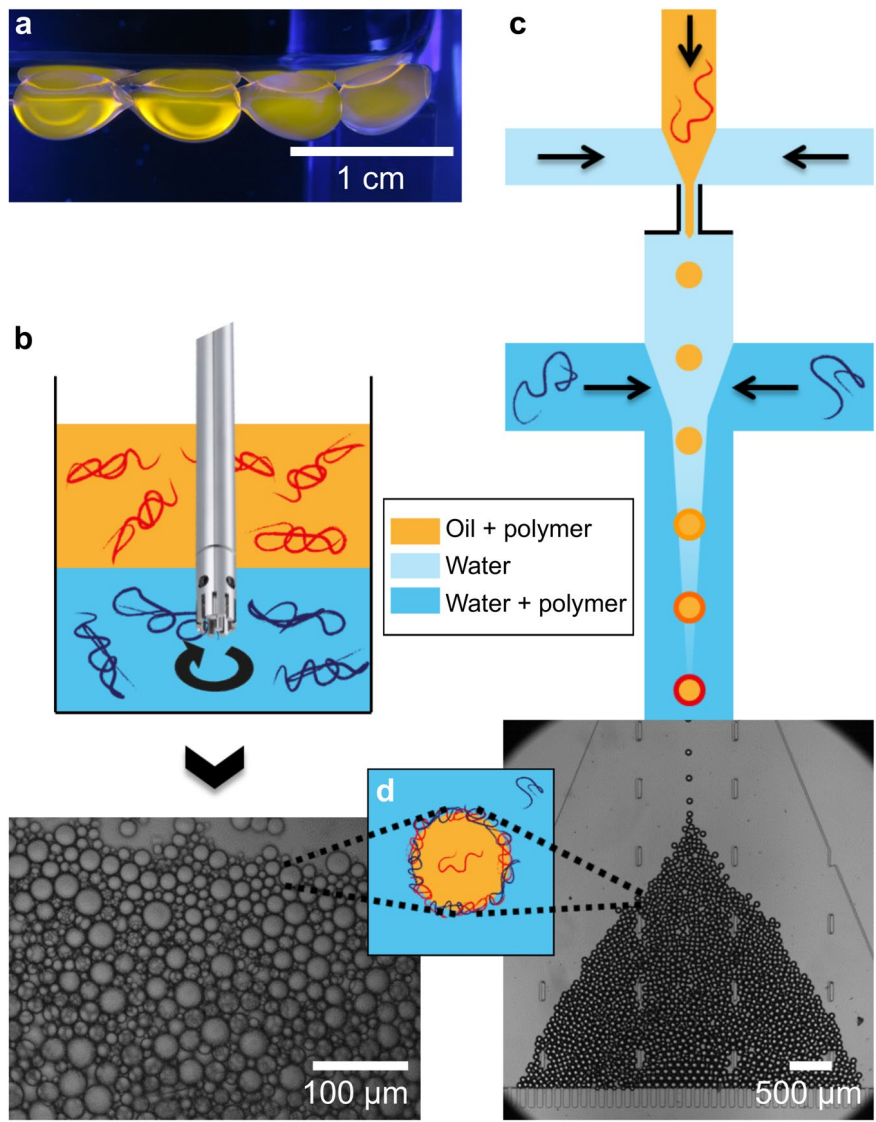

**Figure 3.** Three ways to build multi-sized capsules that remain stable over time. (**a**) Centimetric capsules made by gentle injection of one phase into the other. (**b**) Top: schematic of emulsification performed by a rotor-stator homogenizer in a centimetric vial. Bottom: microscope observation of the obtained emulsion containing polydisperse micron-sized capsules. (**c**) Top: schematic of the microfluidic device used to form capsules in a single step. Oil with PPO is injected in pure water through a flow-focusing unit which forms monodisperse droplets. Water with PMAA is injected after the constriction to avoid early complexation. PMAA diffuses to the water-drop interface, thus leading to the membrane formation. Bottom: microscope observation of the capsules trapped in the chamber by a filter. (**d**) Schematic of a capsule whose membrane formed by interfacial complexation is stable over time.

We thus demonstrated the fine pH-responsiveness of the membrane at moderate pH values. Therefore, pH-triggering can be used to release the oil from the capsules. Evidence for this effect was actually given in our microfluidic experiments. By increasing the pH of the capsules environment from acidic to basic values, the membrane protection vanishes in a few seconds and all the drops coalesce and flow across the filter (Fig. 4(b)).

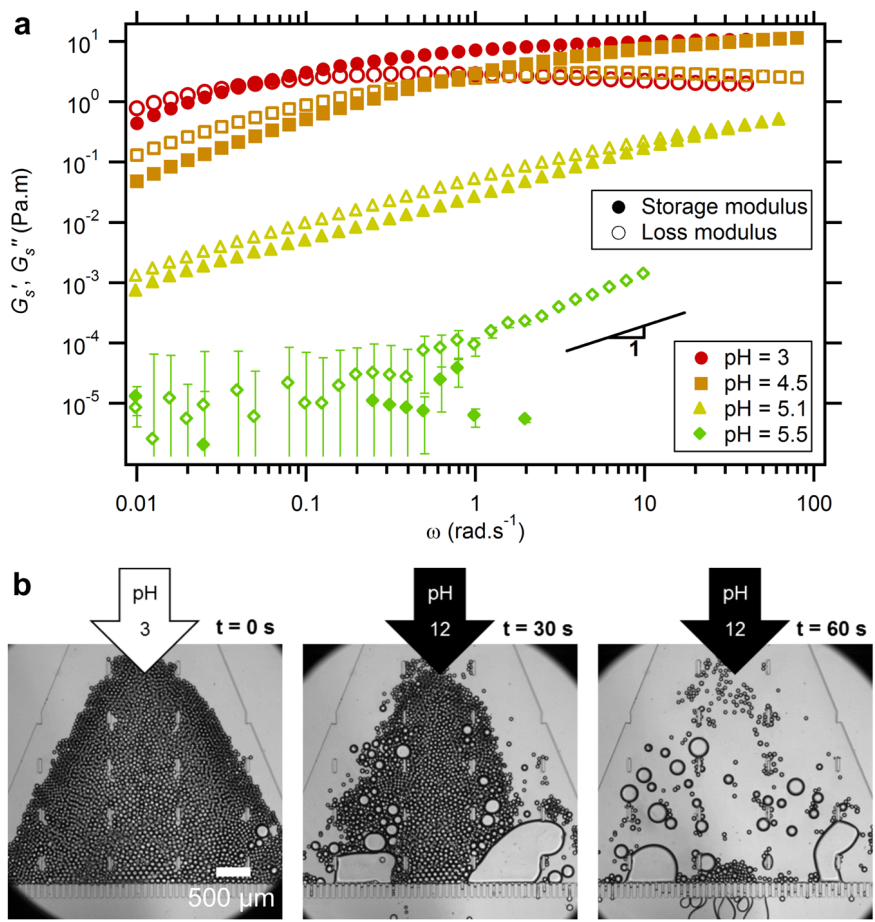

**Figure 4.** (**a**) Frequency sweeps (at strain amplitude = 0.1%) performed on a PMAA-PPO membrane, at several pH from 3 to 5.5. The surfacic storage modulus $G_s'$ is plotted with filled symbols, and the loss modulus $G_s''$ with open symbols. Error bars are shown when larger than markers size. pH increase induces a softening of the membrane. (**b**) Left: capsules stored at pH = 3. Center and right: pH = 12 water flow induces membrane vanishing and inner fluid release.

## Conclusion

We showed that the H-bond complexation of two polymers at liquid interface, one dissolved in oil and the other in water, enables the large production of highly-stable multi-sized biocompatible oil capsules in water. The obtained interfacial membrane exhibits striking self-healing properties. Indeed, after applying external damaging, the polymers quickly diffuse to the interface and reassemble into a rejuvenated structure. H-bond interfacial polymer assembly is a fast and spontaneous process, which can easily be used in different one-step methods to reach large scale production of capsules with controlled size ranging from micrometric to centimetric. At mild values, pH is an efficient and precise stimulus, not only to control the mechanical properties of the membrane form hard to soft materials, but also to dissolve it to release the encapsulated oil. We anticipate that this work will offer a new perspective to targeted and controlled delivery of actives ingredients.

## Methods

Water-based solution is prepared by dissolution of 1 wt% of poly(methacrylic acid) (molar mass: 100000 g.mol$^{-1}$) (Polysciences, Inc.) in water-distilled and purified with milli-Q apparatus (Millipore). Molar mass of a repeat unit MAA is 87.1 g.mol$^{-1}$, which corresponds to a molar concentration of 0.11 mol.L$^{-1}$. pH is adjusted by adding HCl (Sigma-Aldrich) solution concentrated at 1 M or NaOH (Sigma-Aldrich) solution at the same concentration and measured with pH-meter pHM 250 ion analyser Meterlab (Radiometer Copenhagen) with a precision of 0.05 pH.

Oil-based solution is prepared by dissolution of 1 wt% of poly(propylene oxide) (molar mass: 4000 g.mol$^{-1}$) (Sigma-Aldrich) in isopropyl myristate (Sigma-Aldrich) or Miglyol 812 N (IMCD France/Sasol). Miglyol is a neutral oil consisting of caprylic/capric triglyceride (C8/C10 chains). Molar mass of a repeat unit PO is 58.1 g.mol$^{-1}$, which corresponds to a molar concentration of 0.15 mol.L$^{-1}$. We choose 1 wt% for both polymers to ensure an excess of polymer in bulk phases with respect to the interface, while being in dilute regime (<3 wt%) to have a low viscosity solution, hence a better sensitivity to interfacial rheology measurements.

Membrane thickness is measured *in situ* with an optical spectrometer V8E (Specim) assembled on an optical microscope (Olympus). The focus is set strictly at the interface where the membrane grows. Thickness is also measured *ex situ* with an optical profilometer Microsurf 3D (Fogale nanotech) by transferring the membrane from the liquid to a glass slide.

Emulsions are prepared in vials by gently pouring 6 mL of water-based solution and then 4 mL of oil-based solution. An Ultra-Turrax disperser (IKA) is used to form an emulsion with a speed of 20000 rpm during 30 seconds.

To probe interfacial rheometry, an AR-G2 rheometer (TA Instruments) is used with a Double-Wall-Ring geometry. Torque measurement resolution is 1 nN.m. The ring-shaped container is half-filled with approximatively 21 mL of water solution until obtaining a flat interface pinned horizontally between the walls' edges, so the meniscus deformation can be neglected. Then, the ring is carefully approached to the interface and precisely placed to keep a flat interface between the wall corner and the diamond-shaped corner of the ring. Finally, the rest of the container is slowly filled with the same volume of oil. Measurements are controlled by TRIOS software (TA Instruments). A strain rate of 0.1% is imposed to ensure that the measurements are all carried out in the linear regime.

The microfluidic device is first designed with a dedicated software (Clewin) and printed on a mask. Then a SU-8 negative photoresist resin (Micro Chem) is spin-coated with the desired thickness on a silicon wafer. Two-layer lithography is used with no intermediate step[25]. The obtained silica mold is filled with PDMS (poly(dimethylsiloxane)) (Sylgard 184, Dow Corning) mixed with 10% (w/w) curing agent and incubated at 70 °C for around 4 hours. The PDMS is peeled off the mold and the entrances and exits are punched with a 0.5 mm-diameter Harris Uni-Core biopsy punch (Electron Microscopy Sciences). The PDMS is finally sealed to a glass slide with a PDC-002 oxygen plasma cleaner (Harrick Plasma). Every entrance or exit of the chip is connected through PEEK (poly(etheretherketone)) tubings to a small vial. Applied pressures in vials (around 400 mbar) are controlled by a MFCS pump (Fluigent).

### Acknowledgements

This study has been supported through the French National Research Agency by the JCJC INTERPOL project (Grant No. ANR-12-JS08–0007), as well as the Global Station for Soft Matter, a project of Global Institution for Collaborative Research and Education at Hokkaido University. We thank IMCD for having generously provided the Miglyol 812 that we used in our experiments. We acknowledge Caroline Fradin for preliminary rheological measurements.

### Author Contributions

M.R. and C.M. developed the concept and supervised the project. J.D.B. and C.T. designed the experiments; J.D.B. performed the experiments and analyzed the data. C.T., T.S., N.P. and P.P. gave technical support and conceptual advices. All authors discussed and interpreted results. J.D.B. prepared the manuscript; J.D.B., P.P., T.S., M.R. and C.M. edited the manuscript; all authors reviewed and approved the manuscript.

### Additional Information

**Competing Interests:** The authors declare that they have no competing interests.

**Publisher's note:** Springer Nature remains neutral with regard to jurisdictional claims in published maps and institutional affiliations.